# Magnetisation dynamics in the normal and condensate phases of UPd$_2$Al$_3$

## I: Surveys in reciprocal space using neutron inelastic scattering

A. Hiess[1], N. Bernhoeft[2], N. Metoki[3], G. H. Lander[3,4], B. Roessli[5],

N. K. Sato[6], N. Aso[7], Y. Haga[3], Y. Koike[3], T. Komatsubara[8], and Y. Onuki[9]

[1] Institut Laue Langevin, BP 156X, F-38042 Grenoble, France

[2] Dépt. de Recherche Fond. sur la Matière Condensée, CEA-Grenoble, F-38054 Grenoble, France

[3] Advanced Science Research Centre, Japan Atomic Energy Research Institute, Tokai, Naka, Ibaraki 319-1111, Japan

[4] European Commission, JRC, Institute for Transuranium Elements, Postfach 2340, D-76125 Karlsruhe, Germany

[5] Laboratory for Neutron Scattering, ETH Zurich and Paul Scherrer Institute, CH-5232 Villigen, Switzerland

[6] Dept. of Physics, Nagoya University, Furo-cho, Chikasu-ku, Nagoya 464–8602, Japan

[7] Neutron Scattering Laboratory, Institute for Solid State Physics, Univ. of Tokyo, Shirakata 106-1, Tokai-mura, Ibaraki 319-1106, Japan

[8] Physics Department, Graduate School of Science, Tohoku University, Sendai 980-8578, Japan

[9] Graduate School of Science, Osaka University, Toyonaka 560–0043, Japan

## Abstract:

This paper, I, presents new results from neutron inelastic scattering experiments on single crystals of UPd$_2$Al$_3$. The focus is on the experimental position whilst the sequel, II, advances theoretical perspectives. We present a detailed and complete characterisation of the wavevector- and energy-dependent magnetisation dynamics in UPd$_2$Al$_3$ as measured by neutron inelastic scattering primarily in the form of extensive surveys in energy-momentum space under a wide range of experimental conditions, and put our observations in context with data that has been previously published by two independent groups. In this way we emphasize the commonality and robust nature of the data which indicate the intricate nature of the dynamic magnetic susceptibility of this material. Our results yield unique insight into the low temperature ground state which exhibits a microscopic coexistence of antiferromagnetism and superconductivity making UPd$_2$Al$_3$ one of the most accessible heavy-fermion superconductors that can be fully characterised by neutron spectroscopy.

## 1. INTRODUCTION

The co-existence of magnetism and superconductivity continues to attract the attention of the condensed matter community. It is of particular interest to establish whether the superconducting state is stabilised via a dynamic deformation of the lattice, magnetic or other electronic potential. For both high-$T_c$ and heavy-fermion superconducting materials the discussion has been, and still is, extremely controversial. Three fundamental questions arise: First, is it meaningful to discuss superconductivity and magnetism as two separate phenomena or are they joint manifestations of a novel low temperature ground state? Second, on assuming some reduction of the two aspects may be made, what are the symmetries of the order parameters and finally, can one identify coupling mechanisms that maintain the broken symmetry of the appropriate wave function?

Of the materials that are known to exhibit both ordered magnetism and superconductivity, the compound $UPd_2Al_3$ has an especially interesting place. Initially investigated by C. Geibel and collaborators[1] it has the following favourable properties. First, a simple atomic structure, hexagonal space group P6/mmm (a = 5.350 Å, c = 4.185 Å), and the possibility to grow stoichiometric, bulk superconducting single crystals of ~ 2-3 g. Second, a simple antiferromagnetic structure, $T_N$ = 14.3 K, with ferromagnetic sheets of uranium moments parallel to [1 0 0] stacked in alternating directions along the hexagonal **c**-axis, see **Fig. 1,** giving an antiferromagnetic wave vector $\mathbf{Q_o}$ = (0 0 1/2) reciprocal lattice units (rlu) [2, 3]. Third, superconductivity coexists with antiferromagnetic order below a relatively high temperature of ~ 1.9 K giving an energy scale accessible to modern high resolution neutron spectrometers. From the large specific heat and concomitant jump at $T_{sc}$ of $\Delta C = 1.2\ \gamma T_{sc}$ ($\gamma$ = 140 mJ/mol-K$^2$) [1] it has been suggested that the superconducting ground state evolves out of interactions between heavy quasiparticles at the Fermi surface. Finally, $UPd_2Al_3$ possesses a set of intriguing physical properties amongst which number, a significant uranium moment ~ 0.85 $\mu_B$ [2, 3, 4], and, below $T_{sc}$ the absence of a Hebel-Slichter peak,[5] a $T^3$ dependence of the nuclear-spin relaxation time, $T_1$ [6], and the power-law behaviour of the specific heat,[7] all of which have prompted suggestions of unconventional superconductivity.

Of the many techniques available to characterise the spectral magnetic response of this system, neutron inelastic scattering is one of the most powerful giving information on the electronic and nuclear dynamics over temporal ($10^{-13}$ to $10^{-10}$s) and spatial (~ 400 Å) scales ideally suited to investigation of both magnetic and superconducting phenomena. A general formalism, based on linear response theory, relating the cross section to the dissipative component of the magnetic susceptibility (Im $\chi$), exists for the scattering of the neutron against stable thermodynamic states.[8] Within its domain of validity, this enables the inference of direct microscopic information on the dynamic evolution of the magnetic quasiparticle–hole excitation spectra in correlated magnetic macrostates. In the superconducting state the response is modified by the dynamical restrictions imposed by the phase correlated condensate.[9] In addition to the contribution from the quasiparticle-hole excitations of the normal state, the neutron may also couple directly to the superconducting ground state via transitions associated with excitation/condensation of Cooper pairs. As with the normal state excitations, the amplitude of the response depends on the space-time symmetry of the condensate, and, in favourable circumstances, one may observe its signatures through its contribution to the magnetic excitation spectrum. As we shall see, in UPd$_2$Al$_3$ this is indeed the case.

Attempts to examine other heavy-fermion superconductors, e.g. UPt$_3$ [Ref. 10], URu$_2$Si$_2$ [Ref. 11, 12], UBe$_{13}$ [Ref. 13], UNi$_2$Al$_3$ [Ref. 14], by neutron inelastic scattering have all been hampered by the difficulty that the dynamic correlations are weak. In the case of ferromagnetic superconductors such as UGe$_2$ [Ref. 15] relevant experiments to access the superconducting ground state would have to be performed under substantial pressures (10 ~ 15 kbars) and low temperatures T$_{sc}$ ~ 0.2 K. Similar temperature restrictions in the recently discovered ambient pressure systems, ZrZn$_2$ [Ref. 16] and URhGe [Ref. 17], make neutron inelastic scattering experiments difficult from the viewpoint of the temperatures needed as well as the extremely high resolution required to access fluctuations on the scale of T$_{sc}$ (~ 20 μeV). These problems are compounded by the intrinsic problem of the separation of nuclear and magnetic contributions to the cross section at the ferromagnetic position. Thus, although inelastic scattering has been observed from these materials it cannot be correlated in a simple manner with the dynamics of the changing thermodynamic macrostates involved.

It is the specific combination of physical properties that make a neutron inelastic scattering investigation of the normal to superconducting transition in UPd$_2$Al$_3$ possible on account of a dominant *quasielastic* contribution to the magnetisation autocorrelation function at low energies. This opens an experimental window, via high resolution neutron inelastic scattering, on the low energy dynamics that play a key role both in the formation of the antiferromagnetic heavy-fermion state and the simultaneous antiferromagnetic-superconducting ground state.

**Résumé of previous work using neutron inelastic scattering:**

The first neutron inelastic scattering work on single crystals was at Risø National Laboratory in which broad excitations with a strong dispersion along the **c*** ([0 0 1]) axis up to ~ 8 meV at the magnetic zone boundary (where the full width half maximum (*fwhm*) is ~ 9 meV) were reported.[18] In the basal plane strongly damped excitations were found, with poles and widths of similar extent, increasing up to ~ 4 meV. These studies, carried out with 0.3 meV resolution (*fwhm*), found no low energy gap in the excitations at the magnetic zone centre, $Q_o$, and no change when the material became superconducting. However, since the energy resolution was on the scale of ~ 3 K, it is perhaps not surprising that no effect was observed below $T_{sc}$.

Work on polycrystalline material at the ISIS spallation source by Krimmel *et al.*[19] then followed giving an overview of the inelastic response function up to ~ 40 meV. This study gives no clear evidence for a discrete crystal field level scheme and the principle results of these experiments were, that (a) over the studied range of wave vectors a broad quasielastic contribution was present in the scattering at all measured temperatures with a *fwhm* of 9.8 meV at T = 25 K and 22.8 meV at 150 K, and (b) at T = 25 K a strong maximum in the scattered intensity with an energy transfer ~ 2.2 meV at $|Q| \sim 1$ Å$^{-1}$ was identified.

Experiments on single crystals were made by the Tohoku University group using the JRR-3M research reactor (JAERI, Tokai) [N. Aso, PhD-thesis, Tohoku Univ. 1996; unpublished] which motivated higher-resolution experiments at the Institut Laue Langevin, Grenoble (ILL) in 1996 [20]. Around this period a parallel effort started by the group at the Advanced Science Research Centre of JAERI in Tokai, Japan.[21–22] Over the following years several papers have been published concentrating on the magnetic response in the vicinity of $Q_o$, the magnetic zone centre, including polarisation analysis, temperature and field dependent studies. This has resulted in a disparate literature, masking rather than highlighting the fundamental importance and remarkable degree of agreement between data collected on different samples by independent experimental groups. A point of much interest has been the exploitation of initial results obtained by Metoki *et al.*[23] with high energy resolution techniques to resolve the significant intensity around a second characteristic wave vector, $Q*$ = (1/2 0 1/2); this aspect, investigated in more detail at the ILL and Paul Scherrer Institute (PSI), led to an alternative perspective on the origin of the $Q_o$ = (0 0 1/2) Bragg peaks[24].

In parallel with the experimental program, theoretical efforts have been underway to understand the rather unusual effects reported. Early approaches by Sato *et al.*[20,25] were followed by those of Bernhoeft *et al.*[26-30] which exploited the changes in wave vector and energy dependencies of the neutron inelastic scattering amplitude below $T_{sc}$ to infer the symmetry of the energy gap in an analysis based on the role of the phase coherence intrinsic to the superconducting macrostate. More recently Sato *et al.*[31] have published an alternative interpretation of the *same* data building on some aspects of the interpretation given in Refs 26-30. Whilst further work[32] on tunnelling into carefully prepared films supports the interpretations drawn in Refs. 26-30, various other conclusions on the energy gap symmetry, together with more general remarks about the potential driving the superconductivity[31-35] have also appeared.

In view of the general interest generated by the data from these experiments, which arises from their rich information content with respect to the superconducting energy gap symmetry and magnitude, further experiments using cold and thermal three-axis spectrometers were recently performed. This paper provides a comprehensive coverage of the current experimental situation. Important new data is presented mainly in the form of extensive surveys in energy-momentum space under a wide range of experimental conditions. All comparable data presented are consistent between experiments performed on independent samples at JAERI, ILL, and PSI.

To avoid confounding the data, which stand alone, with interpretations, the analytic reduction of the results is deferred to Part II wherein the focus is on the phase coherence and lattice periodicity symmetries and constraints which need to be respected in any given approach. It is hoped that the combination of papers, I and II, may stimulate an interaction between theoretical modelling and further experiments in developing an understanding of the antiferromagnetic superconducting state.

## 2.     EXPERIMENTAL RESULTS

The experiments have been performed on two different samples at JEARI, and the ILL and PSI, respectively. The crystals, with a nominal composition of $UPd_{2.02}Al_{3.03}$, were grown from a melt of high purity elements by the Czochralski method[37,38]. They have a typical mass ~ 2.5 g, are cylindrical in shape and show a mosaic spread of about 1 degree. Both samples exhibit a superconducting transition at ~ 1.9 K.

## A     Overview of the effects around the magnetic zone centre $Q_o = (0\ 0\ 1/2)$

Data on the magnetisation dynamics at, and close to, the magnetic zone centre along the *c** direction is given in Fig. 2. In panel (a) an overview of the temperature evolution at $Q_0$ is afforded. At high temperatures the response is, within the experimental energy resolution of 0.09 meV (*fwhm*), quasielastic. Between $T_{sc}$ and ~ $T_N$/2 the quasielastic energy linewidth at constant intensity scales approximately with $k_BT$ indicating the susceptibility to be more or less temperature independent. In addition to this low energy response, on cooling a distinct, all be it broad, inelastic feature (green arc), nominated as a spin wave[18] or exciton mode[31] appears. At $Q_o$, for temperatures below 2.5 K, this latter mode is observed at an energy transfer E ~ $k_BT_N$ (1.5 meV) with a width of ~ 0.4meV (*fwhm*). This feature remains unchanged when passing into the superconducting phase down to the lowest temperatures measured, 0.15 K. In contrast, at and below 1 K a patent change in profile below 1.4 meV occurs with the quasielastic response being replaced by a distinct excitation (diffuse orange area in lower left hand side of panel (a)), which is characteristic of the low temperature, T < $T_{sc}$/2, superconducting state. For all temperatures studied this latter pole has little change in amplitude, width or position below $T_{sc}$/2 in accordance with it being a quantum excitation.[30]

The dispersions of both the low and high energy features parallel to the hexagonal axis in the vicinity of $Q_o$, are given in **Fig. 2 (b)** at T ~ $T_{sc}$. The former has a quasielastic lineshape at this temperature and exhibits a marked decay in amplitude with increasing wave vector along $q_l$ whilst the latter, inelastic mode, retains its amplitude and form. In **Fig. 2 (d)** the dispersions are given for T << $T_{sc}$. Both the spin wave and condensate modes have an inelastic lineshape which is maintained on moving away from $Q_o$ with, once again, a collapse of the low frequency amplitude. Finally, in panel **2(c)**, the dispersion around $Q_o$ as measured in the (1 0 0) zone is given at 0.15 K. Comparison with the data in the first Brillouin zone is given by the solid line through the (1 0 1/2) data points which represents a smooth fit to the scan at (0 0 1/2) of panel (d) reduced by the factor 2.2 (with a constant background subtracted). The fall off with |**Q**| is consistent with the known uranium (elastic) form factor and shows that *both* the condensate and spin wave poles arise from a magnetic density of similar spatial extent.

At an energy transfer of 0.4 meV, corresponding with the peak of the response in the superconducting phase, the relative normal and superfluid state spatial extent of the magnetic correlations along *c\** may be inferred from the *q*-scans shown in **Fig. 3**. The widths in the normal and superconducting states correspond to a length scale in real space of ~ 100 Å. Thus one surmises that the slow (~ $10^{-11}$s) antiferromagnetic correlations, which change strongly on passing below $T_{sc}$, arise from regions of ~ 100 Å in extent along the hexagonal axis in both the normal and superconducting phases. This figure also demonstrates that, both above and below $T_{sc}$, at the given energy transfer, there is no other response away from $\mathbf{Q_o}$ in the *c\** direction, see also Fig. 6 below.

At comparable temperatures the thermal evolution in the normal state in the hexagonal plane ($q_h$ 0 1/2) and along the hexagonal axis (0 0 $q_l$)  are given in the left and right hand panels of **Fig. 4**. The top panels, at T ~ $T_N$, indicate an anisotropic quasielastic response with the correlations along the (antiferromagnetic) hexagonal axis being of shorter range than those in the (ferromagnetic aligned) hexagonal planes. The lower panels, for T < $T_N$, show a striking fall off in quasielastic intensity on moving away from $\mathbf{Q_o}$, whilst the inelastic (spin wave or exciton) feature around 1.5 meV continues unabated to at least $q_h$ ~ 0.08 rlu with a weak dispersion.

Given the strong, qualitative, change in character of the low energy response on passing below $T_{sc}$ it is important to establish the nature of the peak occurring for T << $T_{sc}$. Careful polarisation analysis has been carried out at $\mathbf{Q_o}$, and we refer to Fig. 2 of our previous publication [Ref. 26] for details. These results establish that the entire dynamical response for 0.15 < T < 10 K is predominately spin reversing (i.e. time asymmetric), *transversely polarised* to the magnetic moment and, taken with the $\mathbf{Q} \times (\mathbf{Q} \times \mathbf{M})$ selection of the neutron dipole cross section, polarised in the hexagonal basal plane. A longitudinal contribution, characteristic of modes polarized parallel to the bulk moment, is not observed below 10 K. These results eliminate scenarios in which the quasielastic response of the normal state is destroyed on entering the superconducting phase and replaced by, for example, a phononic contribution.

The neutron inelastic scattering spectra of Figs. 2 and 3 clearly highlight the change in the low energy response on passing below $T_{sc}$ with the *inelastic* signature of the superconducting state being *qualitatively* different from the quasielastic signal in the normal state. Thus, whilst the slow correlations are approximately constant in spatial extent, remain transversely polarised, and strongly focused around $\mathbf{Q_o}$, the internal dynamics rearrange with the evolution of an excitation gap in the magnetic response. However, in contrast with the thermally excited quasielastic scattering of the normal state, this emergent, inelastic response lies significantly above $k_B T$, indicative of quantum excitation. It exhibits a strong dispersion in the vicinity of $\mathbf{Q_o}$ both along the hexagonal axis and in the basal plane.

That the low energy inelastic feature is related to the superfluid state is substantiated by its progressive quenching both on heating at zero field, Figs. 2 (a), and, at T = 0.4 K, under an applied magnetic field.[22] The collapse in both pole position and intensity of the low energy inelastic feature around $B_{c2}$ (= 3.6 T) is strong support for the origin of the low energy inelastic signal being the excitation of quasiparticles out of the paired superconducting ground state.

Despite the dramatic changes in the low-energy excitation spectrum in the superconducting phase when below ~$T_{sc}/2$, the peak at 1.5 meV transfer differs little in its presentation from that in the normal state, see Fig. 2 (b, d). Inferences based on the thermal evolution of this feature depend critically on the assumptions in a given scenario and can lead to quantitative changes in the inferred energy pole and width below $T_{sc}$.[31] Such details, however, are not robust features of the data analysis. They depend *sensitively* upon the modelling and are crucially dependent on the fact that all features both above and below $T_{sc}$ are on the scale of the experimental resolution in $q$. Any meaningful parameterisation must include the evident dispersion and fit all data under a given thermodynamic condition simultaneously.

To summarise, the magnetic response close to $\mathbf{Q_o}$ comprises: (i) a quasi-elastic response or a very low energy pole which are indistinguishable within the available energy resolution at all temperatures T > ~ $T_{sc}/2$, (ii) an inelastic (spinwave or exciton) response in both the normal and superconducting antiferromagnetically ordered states T < $T_N$ and (iii) the dramatic growth of a dominant inelastic feature at energies ~ 0.4 meV in the superconducting phase temperatures below $T_{sc}/2$.

## B    Overview of effects across the Brillouin zone:

The response across the Brillouin zone  well below $T_{sc}$ for energy transfers up to 4 meV is shown in **Fig. 5**. The strong localisation of scattering around $\mathbf{Q_o}$ is evident and the left hand and central panels illustrate the dispersion of the spinwave-like excitation in the (0 0 $q_l$) and ($q_h$ 0 1/2) directions respectively. In contrast, as shown in the central and right hand panels, the response in the basal plane ($q_h$ 0 1/2) is complex in form having a subsidiary maximum at the position $\mathbf{Q^*}$ = (1/2 0 1/2) as first reported by Metoki *et al.*[23] The broad intensity maximum is associated with a range of wavevectors around $\mathbf{Q^*}$ and appears at an energy transfer of ~ 3 meV.

This is further illustrated in the right hand panel of **Fig. 6** for data taken at 1.5 K. Under the same conditions, the left hand panel illustrates the dispersion, increase in width and decay into weak diffuse scattering of the spin wave excitation in the (0 0 $q_l$) direction. Discussion of the spectral response in the normal antiferromagnetic and paramagnetic states is differed to Fig. 9.

The nature of the scattering close to $\mathbf{Q_o}$ is such that, without modelling the magnetic response function, the separation of quasielastic and propagating components is not without ambiguity, see II. On the other hand, significantly away from $\mathbf{Q_0}$ the dispersive spinwave-like mode decays into weak space-time correlations of feeble amplitude. As a model independent approach, which implicitly ignores all coupling and damping effects, **Fig. 7** gives the intensity maxima in the intermediate $q$-region as observed at T = 2 K in the form of a dispersion relation. At small values of $q_l$ and $q_h$ the energy spectra have two distinct components are resolved (given as open circles), whilst above ~ 0.05 rlu away from $\mathbf{Q_o}$, as indicated by the filled circles, the quasielastic response collapses leaving a distinct dispersive mode which has a stiffness differing by ~ 50% in the two directions.

**Figure 8** emphasizes the difference in response at 0.2 K between $\mathbf{Q_o}$ and $\mathbf{Q^*}$ in the superconducting state: at $\mathbf{Q^*}$ there is no observable change in response on entering the superconducting state in sharp contrast with the strong time correlations around $\mathbf{Q_o}$. As Fig. 8 indicates, the response at $\mathbf{Q^*}$ is little influenced by the quantum correlations induced by the coherence of the superfluid state. The absence of an active role of the superconducting phase coherence for excitations of high energy c.f. $T_{sc}$ is not unexpected[9] and has already been noted for the 1.5 meV response at $\mathbf{Q_o}$ (open circles). On short time scales the dynamic magnetic correlations are expected to average out the slow phase-field coherence associated with the pairing potential of the condensate. Furthermore, the absence of an emerging condensate response at $\mathbf{Q^*}$ below $T_{sc}$ underscores both the stability of the antiferromagnetic correlations in the phase coherent state and the axial gap symmetry along $\boldsymbol{c^*}$ [26-30, 36].

Previous work found the cross section for modes propagating in the basal plane to be poorly defined in momentum and energy transfer at all temperatures below $T_N$.[18-23] The thermal evolution in the normal state of the enhanced, broad response at $\mathbf{Q^*}$, which has a typical energy scale of 35 K, is shown in the contour plots of **Fig. 9** at 2.5, 12 and 20 K. Above $T_N$, at 25K, an earlier report by Krimmel *et al.*[19] using a time of flight technique and polycrystalline material noted a similar enhanced response for $|\mathbf{Q}| \sim |\mathbf{Q^*}| \sim 1$ Å$^{-1}$ with a typical energy $\sim 2.2$ meV and width 0.75 meV (*fwhm*). There is no equivalent enhancement around $\mathbf{Q_o}$ at these elevated temperatures as also illustrated in Fig. 4. In **Fig. 10** the left hand panel extends the observations to higher temperatures and energy transfer. In the paramagnetic regime a broad quasielastic response is present at both $\mathbf{Q_o}$ and $\mathbf{Q^*}$. In contrast with the situation at $\mathbf{Q_o}$, at and below $\sim 20$ K, i.e. well above $T_N$, the signal at $\mathbf{Q}^*$ already becomes inelastic with a maximum at about 3 meV. Additionally, as Fig. 8 and the right hand panel in Fig. 10 show, at and below $T_N$ for the smallest energy transfer measured, $\sim 0.2$ meV, there is an enhancement at $\mathbf{Q_o}$ of long time correlations with no similar signal at $\mathbf{Q^*}$ even in the neighbourhood of $T_N$. Thus, for temperatures from 16 K (T > $T_N$) to 0.15 K, the intensity at the $\mathbf{Q^*}$ position has no quasielastic term and exhibits no observable change as the temperature is lowered through $T_{sc}$. This lack of a low energy response in the vicinity of $\mathbf{Q^*}$ in UPd$_2$Al$_3$ may be contrasted with the case of UNi$_2$Al$_3$ which orders at an incommensurate wave vector close to $\mathbf{Q^*}$ at (1/2±0.11 0 1/2) [14, 39].

In summary, at $\mathbf{Q^*}$ and low temperature, there is an inelastic response broad both in energy and wave vector transfer which becomes quasielastic for temperatures above 20 K. However, there is no quasielastic feature at any temperature in the normal antiferromagnetic phase or low energy inelastic response analogous to that seen around $\mathbf{Q_o}$ at temperatures well below $T_{sc}$.

## 3. CONCLUSION

Previous publications[20-31] have concentrated on the response around $\mathbf{Q_o}$ and its temperature dependence. This has also been the main focus of theoretical efforts[24-31,33-35] The present paper affords new insights by extensive and detailed mapping through the Brillouin zone of the temperature dependent response from well below $T_{sc}$ to ~5$T_N$. The maps, which encompass the two major symmetry directions of the reciprocal lattice, parallel and perpendicular to the hexagonal axis, show the magnetic response in $UPd_2Al_3$ is strongly structured both in momentum and energy.

In addition to the rich energy structured response in the vicinity of $\mathbf{Q_o}$ which we have discussed in Refs. 20-30, there is a secondary maximum at the wave vector $\mathbf{Q^*}$ which persists from 150 mK in the antiferromagnetic-superconducting state to above $T_N$ in the paramagnetic phase as illustrated in Figs. 5-10. Whilst the detailed implication on thermodynamic properties of having such multiple wave vector maxima remains unclear they appear as a common theme in strongly correlated electronic systems[10-14] and in the present case it has been proposed that the nominally ordered state in $UPd_2Al_3$ remains dynamic in nature on account of the $\mathbf{Q^*}$ mode. In this respect, the contrasting $q$-space local response at $\mathbf{Q_o}$ and the $q$-space spread form at $\mathbf{Q^*}$, identified in the present studies, may have a more general bearing on the existence of an antiferromagnetic-superconducting ground state.[24]

The task of understanding a coherent antiferromagnetic-superconducting ground state remains a major challenge in condensed matter physics. In the interim, we hope the rich and robust nature of the data on $UPd_2Al_3$ presented here will stimulate further experiments and discoveries of other model systems. In Part II we complement these studies with a critical appraisal of the assumptions, scope and limits inherent in analyses of inelastic neutron scattering data and the modelling of the magnetic response function.

## 4.    ACKNOWLEDGEMENTS

We thank all colleagues who have helped this work, in particular the critique of E. Blackburn, A. Kreyssig and O. Stockert is appreciated. Both GHL and NB would like to thank the Director and staff of the Advanced Science Research Centre, JAERI, for warm hospitality during visits that have advanced this collaboration. AH thanks colleagues at IFP, Technische Universität, Dresden, for hospitality during his visit. The neutron-inelastic scattering experiments have been performed at the Institut Laue Langevin (France), JAERI (Japan) and the SINQ source at PSI (Switzerland).

## **Figure Captions**

**Fig 1: (colour online)** The crystallographic and magnetic structure of UPd$_2$Al$_3$. The large circles represent the positions of uranium ions with the bold arrows marking the relative directions of the magnetic moments. The smaller red circles in the same planes represent the positions of the palladium ions whilst the smallest blue circles, in the intercalating plane, represent the aluminium ions.

**Fig 2: (colour online)** (a) Contour plot of the intensity at $\mathbf{Q_o} = (0\ 0\ 1/2)$ as a function of temperature and energy transfer. Marked on the plot are the energies of the characteristic temperatures $T_{sc}$ and $T_N$ and the line $E = k_B T$ to indicate the approximate division between thermal and quantum induced fluctuations. (b) Dispersion of the inelastic response for $\mathbf{Q} = (0\ 0\ q_l)$ with (from bottom to top) $q_l = 0.500, 0.515, 0.530, 0.545$ at 1.8 K, i.e. close to $T_{sc}$. Note logarithmic vertical scale and the zero level of successive scans are displaced by 1 decade for clarity. (c) Dispersion of the inelastic response for $\mathbf{Q} = (1\ 0\ q_l)$ at 0.15 K. The solid line shown in the $(1\ 0\ 1/2)$ scan is smooth fit to the scan at $(0\ 0\ 1/2)$ of panel (d) reduced by the factor 2.2 (with a constant background subtracted) as expected from the uranium form factor. (d) Dispersion of the inelastic response for $\mathbf{Q} = (0\ 0\ q_l)$ at 0.15 K. The horizontal bar indicates the instrumental resolution. Except when indicated, the statistical error corresponds to the size of the symbols. Data taken at ILL on IN14 with $k_f = 1.15$ Å$^{-1}$. A selected fraction of experimental data used to compile this figure has been previously published [26-31].

**Fig. 3:** Neutron intensities as a function of $q_l$ along the $\boldsymbol{c^*}$ axis taken at constant energy transfer of 0.4 meV in both the normal and superconducting states. The widths correspond to a correlation length in real space of ~ 100 Å. At both 0.45 K and 2 K the response is confined to the immediate vicinity of $\mathbf{Q_o}$. Data taken at JAERI with incident wave vector fixed at $k_i = 1.5$ Å$^{-1}$ with a corresponding energy resolution of ~ 0.2 meV (*fwhm*).

**Fig. 4:** Comparison of the scattering in the hexagonal plane ($q_h\ 0\ 1/2$) (left) and along the hexagonal $\boldsymbol{c^*}$ axis ($0\ 0\ q_l$) (right) at different temperatures in the normal state. Note logarithmic vertical scale and different steps in reciprocal space (a reciprocal lattice unit corresponds to $a^* = 1.355$ Å$^{-1}$ and $c^* = 1.500$ Å$^{-1}$ along the two axes). Only for temperatures ~ $T_N$ does the quasielastic response become more isotropic and extend significantly out into the zone in both the basal plane and along the hexagonal axis. Data taken at ILL on IN14 with $k_f = 1.15$Å$^{-1}$.

**Fig. 5: (colour online)** Contour map at 0.15 K showing the response at relatively low energy transfer across the Brillouin zone. The magnetic zone centres ($\mathbf{Q_o}$) are $(0\ 0\ 1/2)$ and $(1\ 0\ 1/2)$. In the $(0\ 0\ q_l)$ direction (left hand panel) the response is centred around $\mathbf{Q_o}$, whereas it is more complex in the $(q_h\ 0\ 1/2)$ direction. This figure shows the secondary maximum in the inelastic response at the position $\mathbf{Q^*} = (1/2\ 0\ 1/2)$. The abscissa are scaled to accommodate the different $a$ and $c$ axis lattice parameters. The colour scheme, designed to highlight the behaviour around $\mathbf{Q^*}$, leads to a saturation close to $\mathbf{Q_o}$ (for details see Fig. 8 on semi-logarithmic scales). The cross section at the smallest energy transfers is inaccessible due to incoherent elastic scattering, and, at $\mathbf{Q_o} = (0\ 0\ 1/2)$, due to the antiferromagnetic Bragg peak. Data taken at ILL on IN14 with $k_f = 1.3$ Å$^{-1}$.

**Fig. 6: (colour online)** Left and right hand panels give contour plots of the intensity at 1.5 K in the (0 0 $q_l$) and ($q_h$ 0 0 ) directions across the zone as determined with $k_f$ = 2.662 Å$^{-1}$ (resolution ~ 1 meV (*fwhm*)). Note the inelastic dispersive feature emanating from the $\mathbf{Q_o}$ = (0 0 1/2) position, and the response at $\mathbf{Q^*}$ = (1/2 0 1/2), which is centered at about 3 meV and extends to ~ 6 meV. Data taken at ILL on IN8.

**Fig. 7:** Plot of intensity maxima from scans with T ~ 2 K presented as a dispersion relation. The closed symbols indicate data from scans in which a single maximum is observed. Open symbols indicate regions where two features are observed in energy scans. Note that for T < T$_{sc}$ the quasielastic response is replaced by a low-lying excitation, but without any effect on the inelastic feature at 1.5 meV. Away from $\mathbf{Q_o}$, the grey area of ± 0.7 meV indicates the region over which the intensity has at least 50% of its peak value. The dashed line corresponds with a stiffness of 14.6 meV·Å in the *c\** direction (left hand panel) and 10.5 meV·Å in the basal plane (right hand panel). The abscissa are scaled to accommodate the different *a*- and *c*-axis lattice parameters.

**Fig. 8:** Constant *q*-scans at $\mathbf{Q_o}$ and $\mathbf{Q^*}$ taken below T$_{sc}$ showing the absence of any features in the response at $\mathbf{Q^*}$ up to an energy transfer of 2 meV. Data taken at ILL on IN14 with $k_f$ = 1.15 Å$^{-1}$.

**Fig. 9: (colour online)** Inelastic response across the zone from (1/2 0 1/2) to (1 0 1/2) at three temperatures. Note the response (green island) at $\mathbf{Q^*}$ persists with approximately constant intensity both in zero point and thermal excitation from low temperature to well above T$_N$. The intensity recorded at (1 0 1/2) marks an equivalent $\mathbf{Q_o}$ position. Data taken at PSI with $k_f$ = 1.5 Å$^{-1}$.

**Fig. 10:** Left hand panels: constant *q*-scans at $\mathbf{Q_o}$ and $\mathbf{Q}^*$ at different temperatures. Above T$_N$, the response persists to at least 8 meV in energy transfer. Quasielastic scattering is present at both positions for T$_N$ < T < 80 K but only persists below T$_N$ at $\mathbf{Q_o}$. Note the maximum at finite energy transfer in both cases below T$_N$. The data below 1 meV have been suppressed since they fall within the (elastic) resolution window of the spectrometer. Data taken on IN8 with $k_f$ = 2.662 Å$^{-1}$. Right hand panel: Temperature dependence of the low energy response at the two positions $\mathbf{Q_o}$ and $\mathbf{Q}^*$ taken at 0.2 meV energy transfer. Note that the response is strongest at T$_N$ (marked by arrow) at $\mathbf{Q_o}$ with negligible temperature dependence at $\mathbf{Q}^*$. Data taken at PSI with $k_f$ = 1.15Å$^{-1}$.

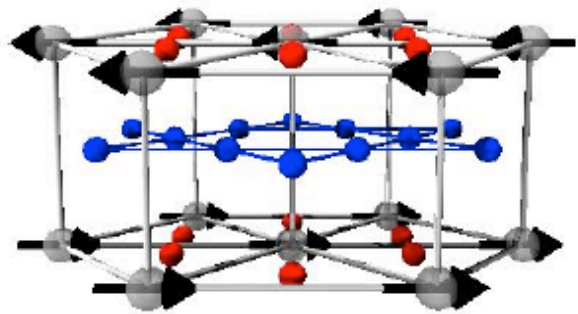

Fig. 1

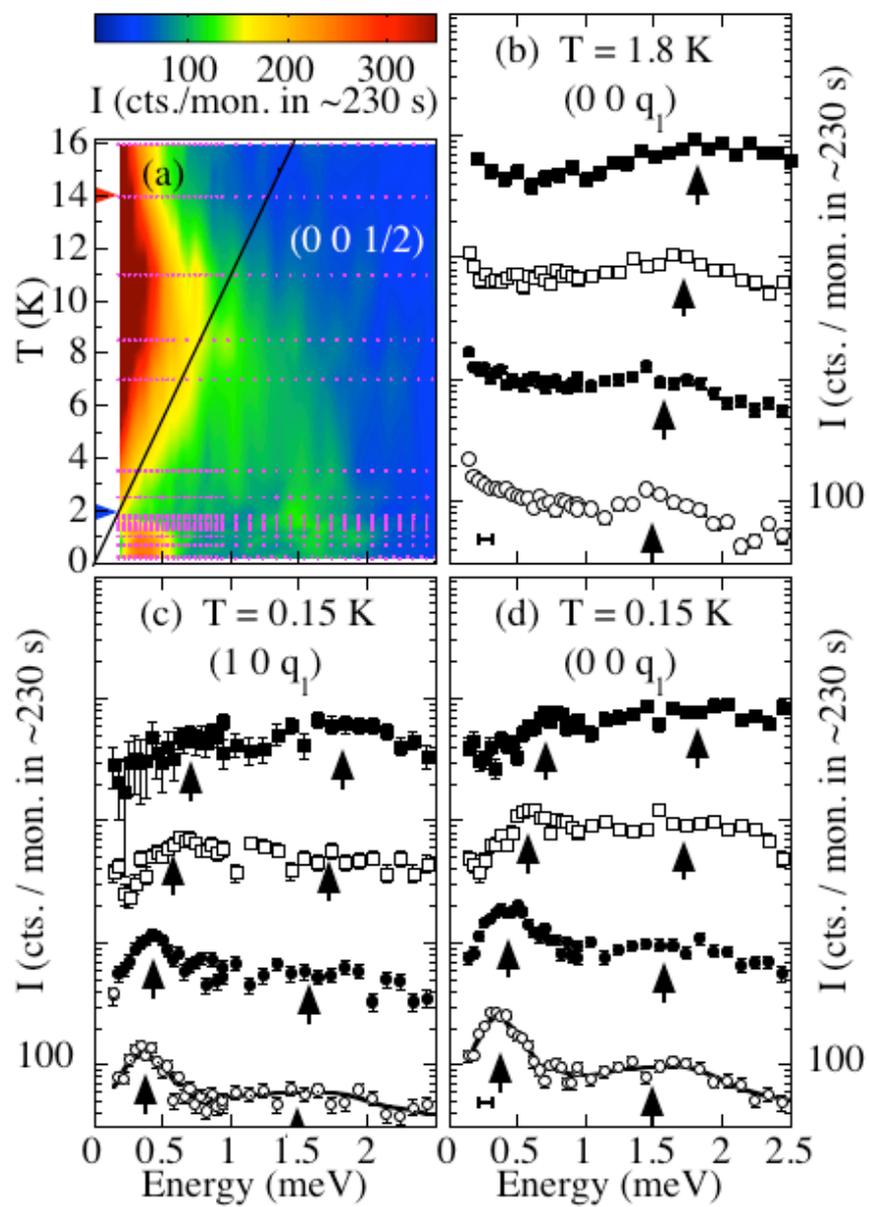

Fig. 2

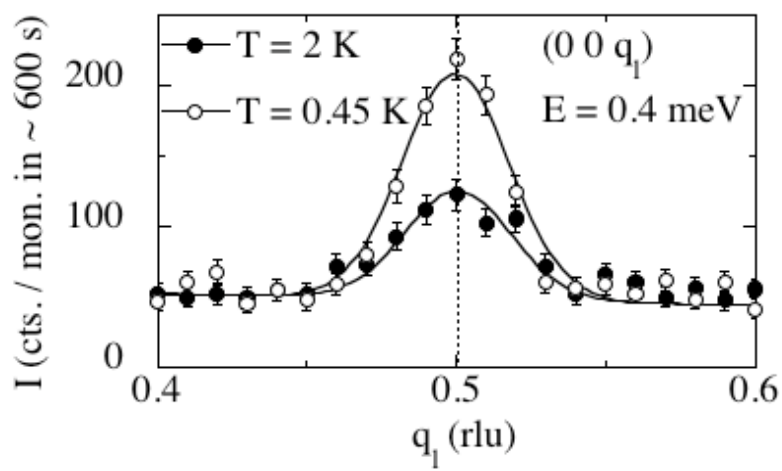

Fig. 3

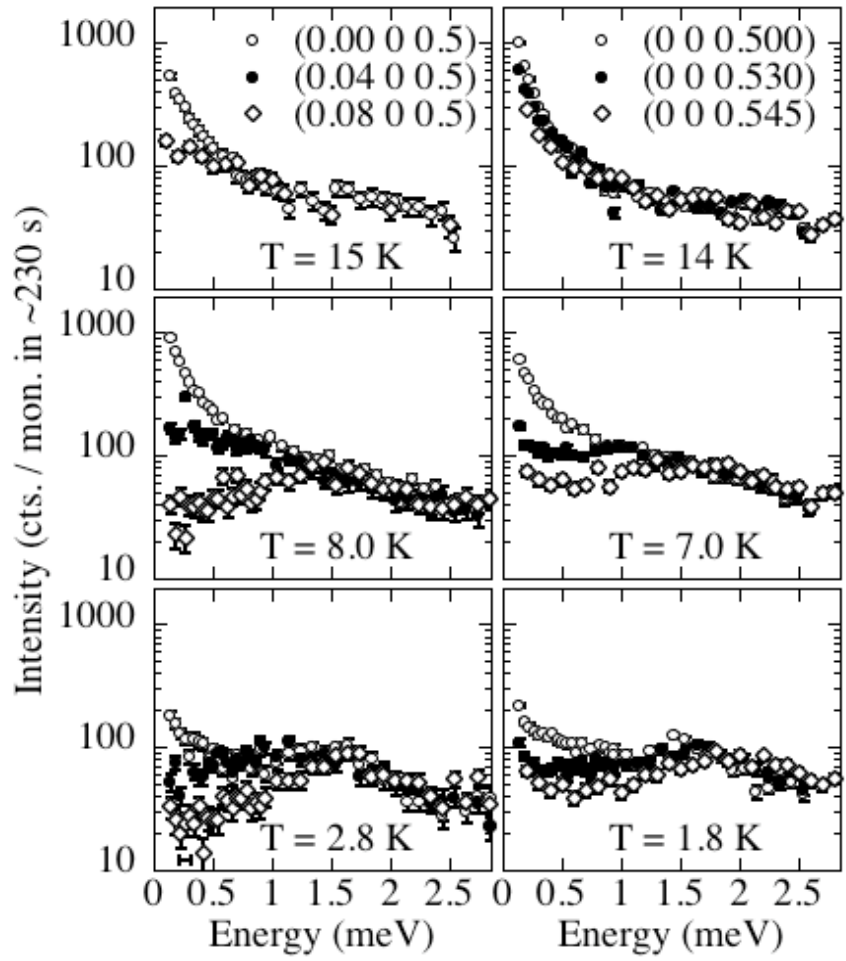

Fig. 4

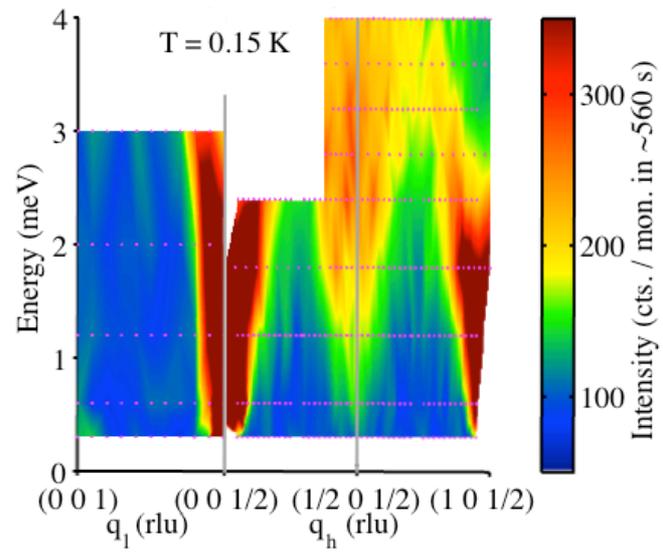

Fig. 5

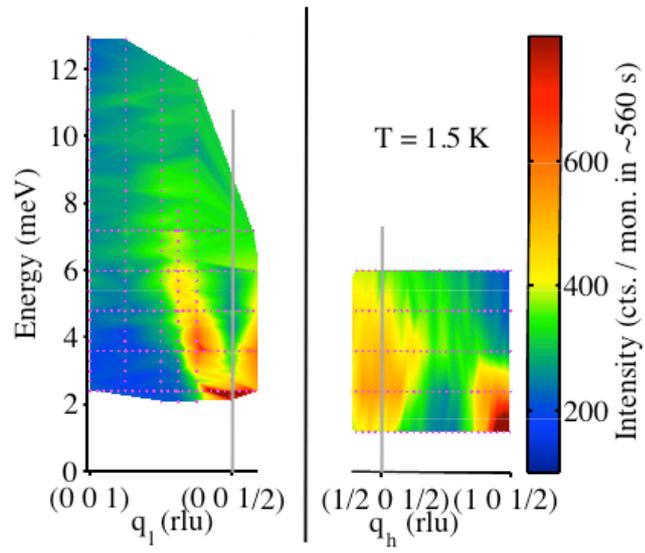

Fig. 6

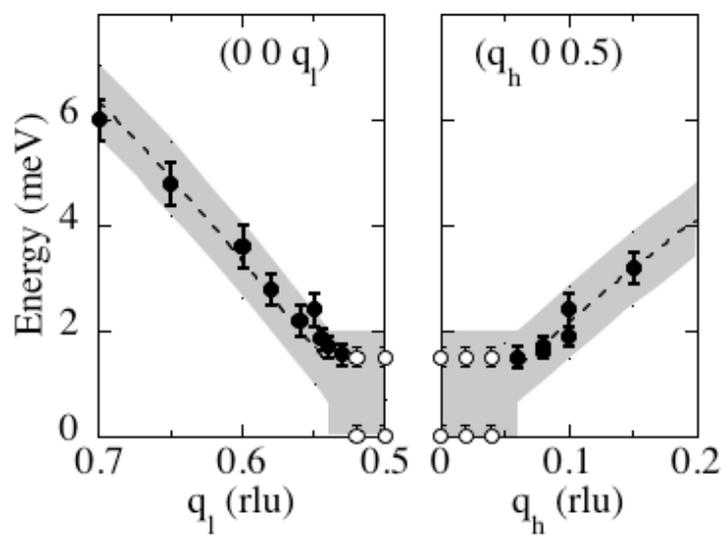

Fig. 7

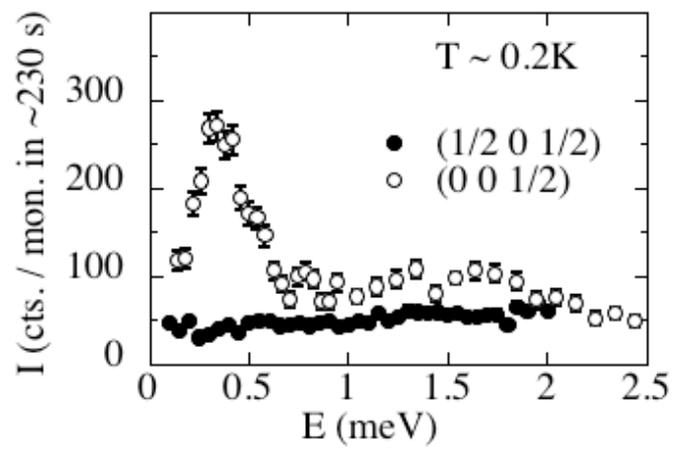

Fig. 8

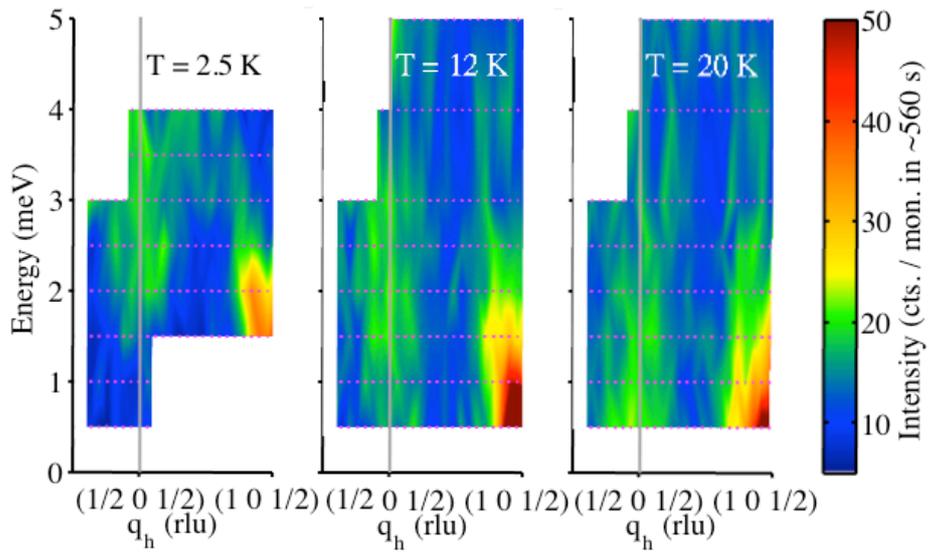

Fig. 9

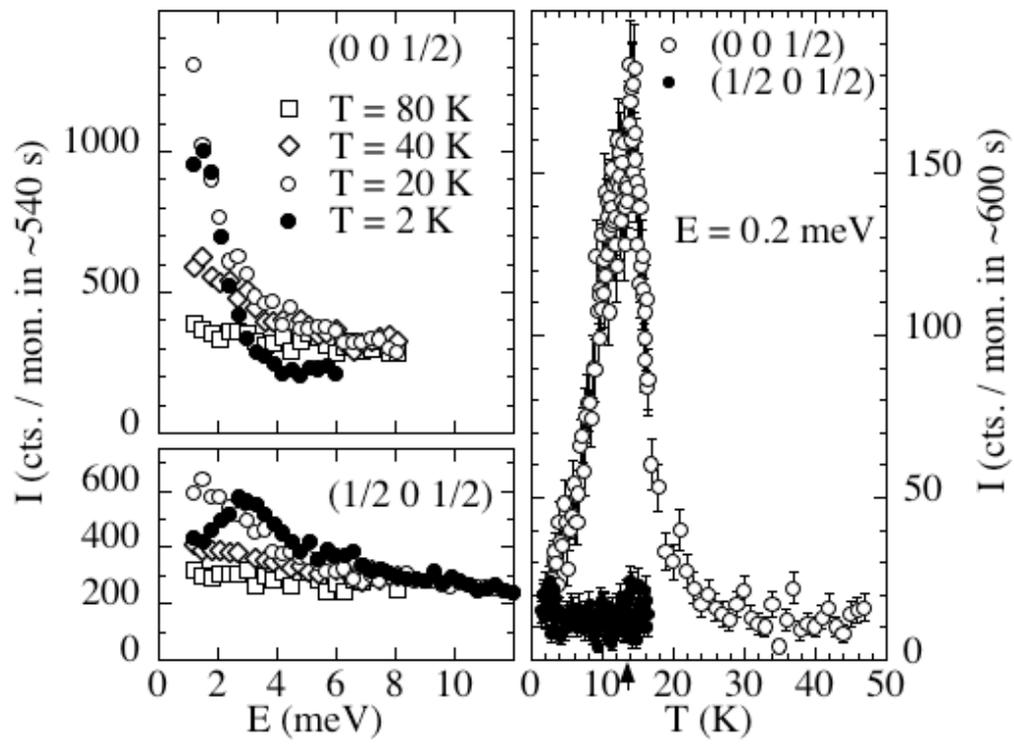

Fig. 10